\documentclass[twocolumn]{aastex62}
\usepackage{graphicx}
\usepackage{dcolumn}
\usepackage{bm}
\usepackage{natbib}
\usepackage{ulem}

\newcommand{\fc}{{\ensuremath{f_\mathrm{\!c}}}}
\newcommand{\fmu}{{\ensuremath{f_{\!\mu}}}}
\newcommand{\Msun}{{\ensuremath{\mathrm{M}_\odot}}}
\newcommand{\cm}{{\ensuremath{\mathrm{cm}}}}
\newcommand{\s}{{\ensuremath{\mathrm{s}}}}

\newcommand{\pcc}{{\ensuremath{\cm^{-3}}}}
\newcommand{\gcc}{{\ensuremath{\mathrm{g}\,\pcc}}}
\newcommand{\kms}{{\ensuremath{\mathrm{km}\,\s^{-1}}}}
\newcommand{\K}{{\ensuremath{\mathrm{K}}}}
\newcommand{\spr}{\textsl{s}-process }
\newcommand{\vrot}{{\ensuremath{v_\mathrm{\!rot}}}}
\newcommand{\vc}{{\ensuremath{v_\mathrm{\!crit}}}}
\newcommand{\vq}{{\ensuremath{v_\mathrm{\!min}}}}

\begin{document} 

\title{New \spr Mechanism in Rapidly-Rotating Massive Pop II Stars}
\correspondingauthor{Projjwal Banerjee}
\email{projjwal.banerjee@gmail.com}
\author{Projjwal Banerjee}
\affil{Department of Physics, Indian Institute of Technology Palakkad, Palakkad, Kerala 678557, India}
\affil{Department of Astronomy,  School of Physics and Astronomy, Shanghai Jiao Tong University, Shanghai 200240, China}
\author{Alexander Heger}
\affil{Monash Centre for Astrophysics, School of Physics and Astronomy,
Monash University, Vic 3800, Australia}
\affil{Tsung-Dao Lee Institute, Shanghai 200240, China}
\author{Yong-Zhong Qian}
\affil{School of Physics and Astronomy, University of Minnesota, Minneapolis, Minnesota 55455}
\affil{Tsung-Dao Lee Institute, Shanghai 200240, China}

\date{\today}

\begin{abstract}
We report a new mechanism for the \spr in rotating massive metal-poor stars. Our models show that above a critical rotation speed, such stars evolve in a quasi-chemically-homogeneous fashion, which gives rise to a prolific \textsl{s}-process. Rotation-induced mixing results in primary production of $^{13}$C, which subsequently makes neutrons via $^{13}\mathrm{C}(\alpha,\mathrm{n})^{16}\mathrm{O}$ during core He burning. Neutron capture can last up to $\sim 10^{13}\,\s$ ($\sim 3\times 10^{5}$~yr) with the peak central neutron density ranging from  $\sim10^7$ to $10^{8}\,\pcc$. Depending on the rotation speed and the mass loss rate, a strong \spr can occur with production of elements up to Bi for progenitors with initial metallicities of $[Z]\lesssim -1.5$. This result suggests that rapidly-rotating massive metal-poor stars are likely the first site for the main \textsl{s}-process. We find that these stars can potentially explain the early onset of the \textsl{s}-process and some of the carbon-enhanced metal-poor (CEMP-\textsl{s} and CEMP-\textsl{r/s}) stars with strong enrichment attributed to the \spr or a mixture of the \textsl{r}-process and the \textsl{s}-process.
\end{abstract}

\section{Introduction}

Nearly all of the solar abundances of elements heavier than the Fe group came from the slow (\textsl{s}) and rapid (\textsl{r}) neutron capture processes, with approximately equal contributions from each. It is well known that the main component of the \textsl{s}-process producing elements up to $^{209}$Bi is associated with low-mass stars of $\sim 1$--$3\, \Msun$ and occurs during their final stage of evolution, the asymptotic giant branch (AGB) phase \citep{straniero1995,gallino1998,arlandini1999}. In addition, massive stars of $\gtrsim 10\, \Msun$, which end their lives in core-collapse supernovae (CCSNe), are traditionally associated with a weak component of the \spr producing elements with mass numbers up to $A\sim 90$. This process mostly occurs during core He burning and requires relatively high initial metallicities ($[Z]\gtrsim -1$) \citep{peters1968,couch1974,lamb1977,raiteri1991,rauscher2002,pignatari2010}. It was shown, however, that a weak \spr can occur even in massive metal-poor stars provided they rotate sufficiently fast initially \citep{pignatari2008}. Further, a recent study showed that non-rotating massive metal-poor stars with enhanced initial C abundances can also produce a weak \spr \citep{bqh_cemp}. In both cases, the low-metallicity weak \spr primarily produces elements up to $A\sim 90$. Another recent study showed that proton ingestion in convective He shells during the late evolution of non-rotating massive metal-poor stars can lead to a neutron-capture process producing a variety of abundance patterns, including those similar to what is attributed to the main \spr \citep{bqh_pingest}.  

In this paper we focus on models of rapidly-rotating massive stars that reach the so-called quasi-chemically-homogeneous (QCH) state following core H burning. In such stars, rotation is fast enough to overcome the composition barrier so that they turn into essentially He stars by the end of core H burning. In these models, stars have very little He left on the surface by the time they reach the pre-collapse stage, and subsequently would produce He-deficient Type Ic SNe (SNe Ic). For this reason, rotating stars that reach the QCH state were studied previously as possible progenitors for broad-line SNe Ic (SNe Ic-BL) associated with long gamma-ray bursts (LGRBs) \citep{scyoon2005,scyoon2006,woosley2006} or superluminous SNe Ic (SLSNe Ic; \citealt{ad2018}). The primary emphasis of those studies, however, was on the connection with LGRB or SLSN-Ic progenitors rather than on the nucleosynthesis. We find that the same models in fact provide a novel mechanism for a rotation-induced \spr in massive metal-poor stars. 

Whereas a rotation-induced \spr in rapidly-rotating massive metal-poor stars was reported previously, stars in those models do not reach the QCH state and only produce mostly elements up to $A\sim 90$ \citep{frischk2012,frischk2016,choplin2018}. We note, however, that \citet{limongi2018} reported an efficient \spr in rotating metal-poor star of $\lesssim 20\, \Msun$ that do not reach the QCH state. In this case, neutron capture is stronger than the typical weak \spr and can easily produce elements up to the second \spr peak at Ba ($A\sim 138$). The third \spr peak at Pb ($A\sim 208$), however, is produced only in the few models for stars of $\lesssim 15\, \Msun$ with $[{\rm Fe/H]}= -2$. In contrast, we find that neutron capture in QCH models is dramatically different and much more efficient, and can easily produce elements up to Pb usually attributed to the main \textsl{s}-process.

\section{Methods}

We study the evolution and nucleosynthesis of rotating stars of $15\,\Msun$ and $25\,\Msun$ using the 1D hydrodynamic code \textsc{Kepler} \citep{weaver1978,rauscher2003}. We focus on progenitors with an initial metallicity of  $[Z]=-3$, corresponding to $[{\rm Fe/H]}\sim -3$ for our adopted composition. The detailed composition is scaled from the solar abundances for Be to Zn and the results of Big Bang nucleosynthesis are used for H to Li. The implementation of rotation including the treatment of transport of angular momentum and mixing is the same as presented in \citet{heger2000a}. Two free parameters \fc{} and \fmu{} control the efficiency of mixing due to rotation by specifying the contribution from this mixing to the diffusion coefficient and the sensitivity of this mixing to composition gradients, respectively. The default values are taken to be $\fc=1/30$ and $\fmu=0.05$.  These values provide a good fit to the observed enhancement of surface nitrogen abundance for rotating stars of $10$--$20\,\Msun$ \citep{gies1992,villamariz2005} with initial rotation speed of $\sim 200$--$300\, \kms$ at solar metallicity \citep{heger2000a}. The effects of magnetic fields resulting from the Taylor-Spruit dynamo \citep{spruit2002} are also taken into account. In particular, the effect of the magnetic torque on angular momentum transport is included as discussed in detail by \citet{heger2005}. 

We explore models with rotation speeds $\vrot$ ranging from $20\,\%$ to $70\,\%$ of the critical breakup speed $\vc$. This range corresponds to equatorial surface rotation velocities of $\sim 200$--$700\, \kms$ for the  $25\, \Msun$ progenitor with $[Z]=-3$ upon reaching the zero-age main sequence. The multi-zone nucleosynthesis is followed from the birth of a star to its death in a CCSN using a large adaptive co-processing network. The network includes isotopes from the proton-drip line to the neutron-drip line for elements up to At (atomic number 85), with a maximum of 4670 isotopes. We use the rates in the detailed list of \citet{rauscher2002}. The CCSN explosion is simulated by driving a piston through the collapsing progenitor with the piston velocity adjusted to obtain the desired explosion energy \citep{weaver1978}. 

Mass loss is important for the evolution of rapidly-rotating massive metal-poor stars, especially during the Wolf-Rayet (WR) stage. For non-WR evolution, we use the mass loss rate from \citet{nieu1990}. We consider a star to be in the WR stage when its effective surface temperature exceeds $10^4\, \K$ and its surface H mass fraction drops below $0.4$. For WR stars, we compute two types of models where the mass loss rates are reduced from those of \cite{hamann1995} by factors of three (M1) and ten (M2), respectively. The former reduction was suggested by \citet{hamann1998} to account for clumping and the latter was suggested by \citet{scyoon2006} for metal-poor WR stars. We use a scaling law $\propto Z^{0.5}$ to account for the metallicity dependence of the mass loss rate, where $Z$ is taken to be the initial metallicity of the star and does not include any heavy elements produced during its evolution.

\begin{figure*}[h]
\centerline{\includegraphics[width=\textwidth]{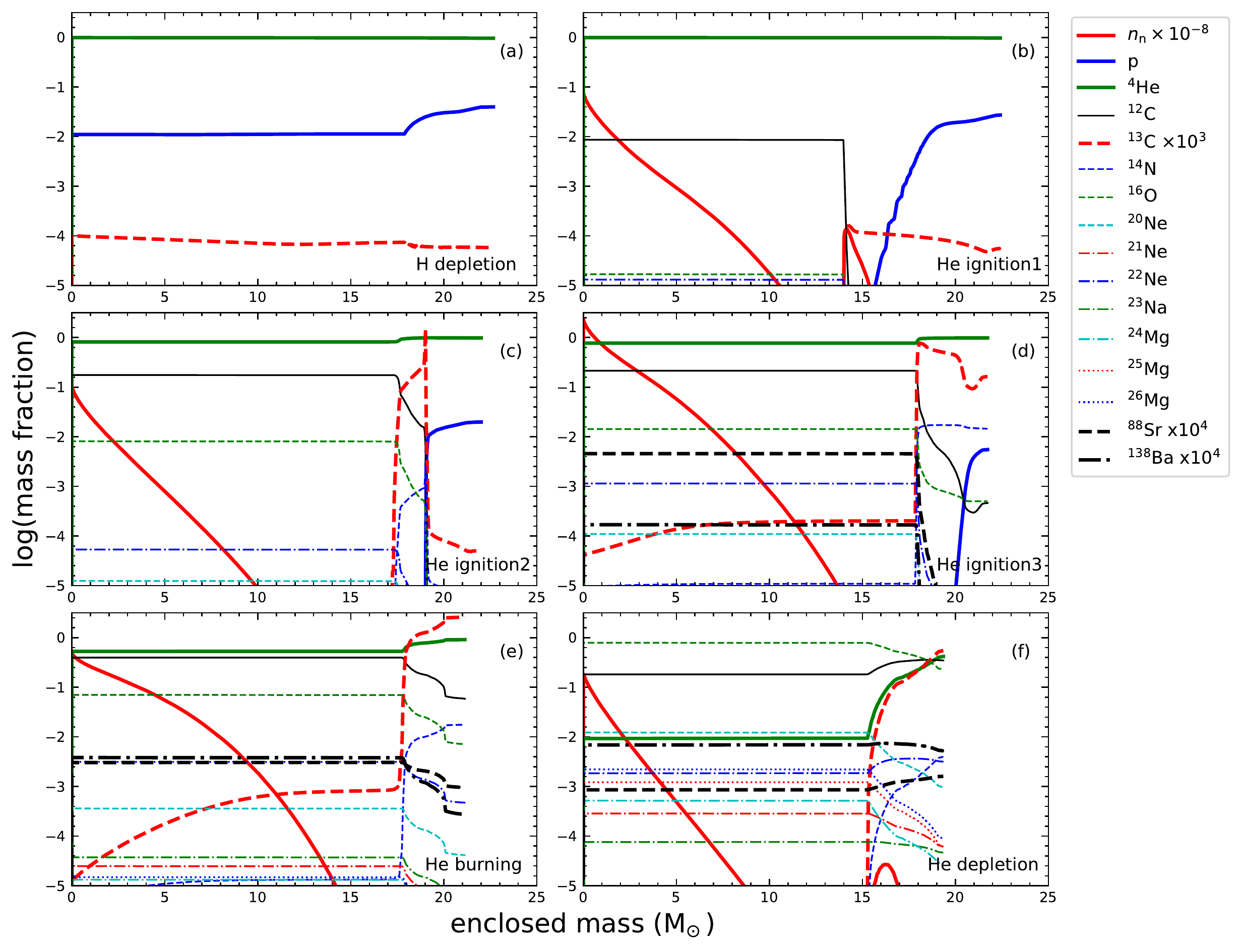}}
\caption{Neutron number abundance (in \pcc) and mass fractions of other light isotopes as functions of mass coordinate at different stages of stellar evolution, from central H depletion (a) to central He depletion (f) for the M1 model of a $25 \,\Msun$ star with an initial metallicity of $[Z]=-3$ and an initial rotation speed of $\vrot=0.5\,\vc$. 
Neutron number abundance is scaled down by a factor of $10^8$.
Mass fractions of $^{13}$C is scaled up by a factor of $10^3$. Mass fractions of the neutron-capture isotopes $^{88}$Sr and $^{138}$Ba are scaled up by a factor of $10^4$ and shown for reference. (a) H depletion: when the central H mass fraction drops to $1\,\%$ following core H burning, (b) He ignition1: when the central $^{12}$C mass fraction reaches $1\,\%$, (c) He ignition2 and (d) He ignition3: intermediate between (b) and (e), (e) He burning: when the central He mass fraction drops to $50\,\%$, (f) He depletion: when the central He mass fraction drops to $1\,\%$.}
\label{fig:snap}
\end{figure*}

\section{Results}
We find that for $\vrot$ above a certain fraction of $\vc$, stars in all our models reach the QCH state following evolution on the main sequence. This result is consistent with previous studies where similar QCH evolution was found \citep{scyoon2005,woosley2006}. Stars in those studies essentially become rigidly rotating He stars with surface H mass fraction of $<10\,\%$ by the time He is ignited at the center. Because of the rigid rotation, the density gradient is shallow, with $\gtrsim95\,\%$ of the star's mass at densities of $\gtrsim 0.1\,\gcc$ and very little mass in a low-density envelope. The minimum initial rotation velocity $\vq$ required for the transition to the QCH state depends slightly on the mass, with lower mass progenitors requiring faster rotation. For example, in models with $[Z]=-3$, this transition occurs at $40\,\%$ and $30\,\%$ of the critical velocity for stars of $15\,\Msun$ and $25\,\Msun$, respectively.

\begin{figure}
\centerline{\includegraphics[width=85mm]{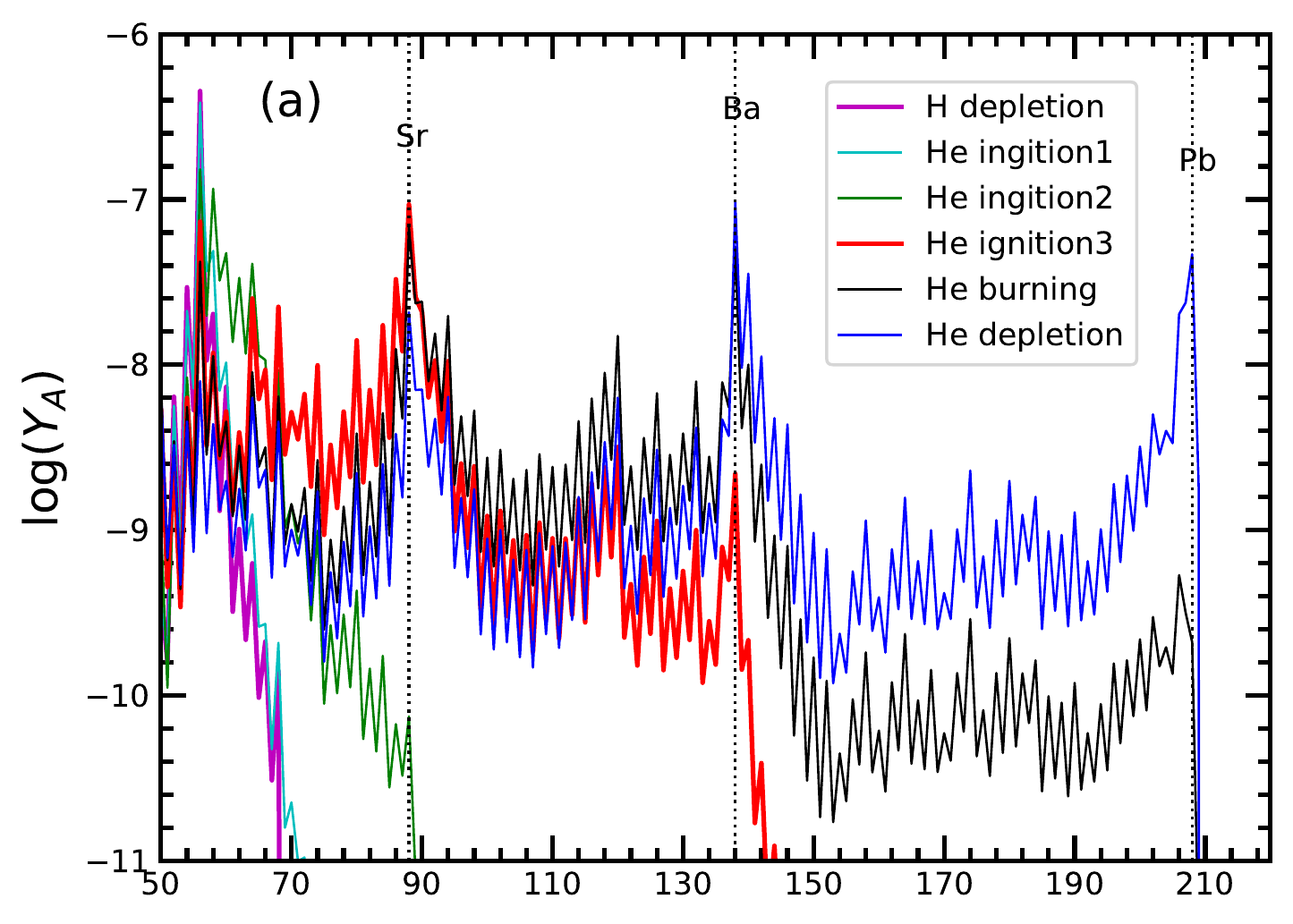}}
\centerline{\includegraphics[width=85mm]{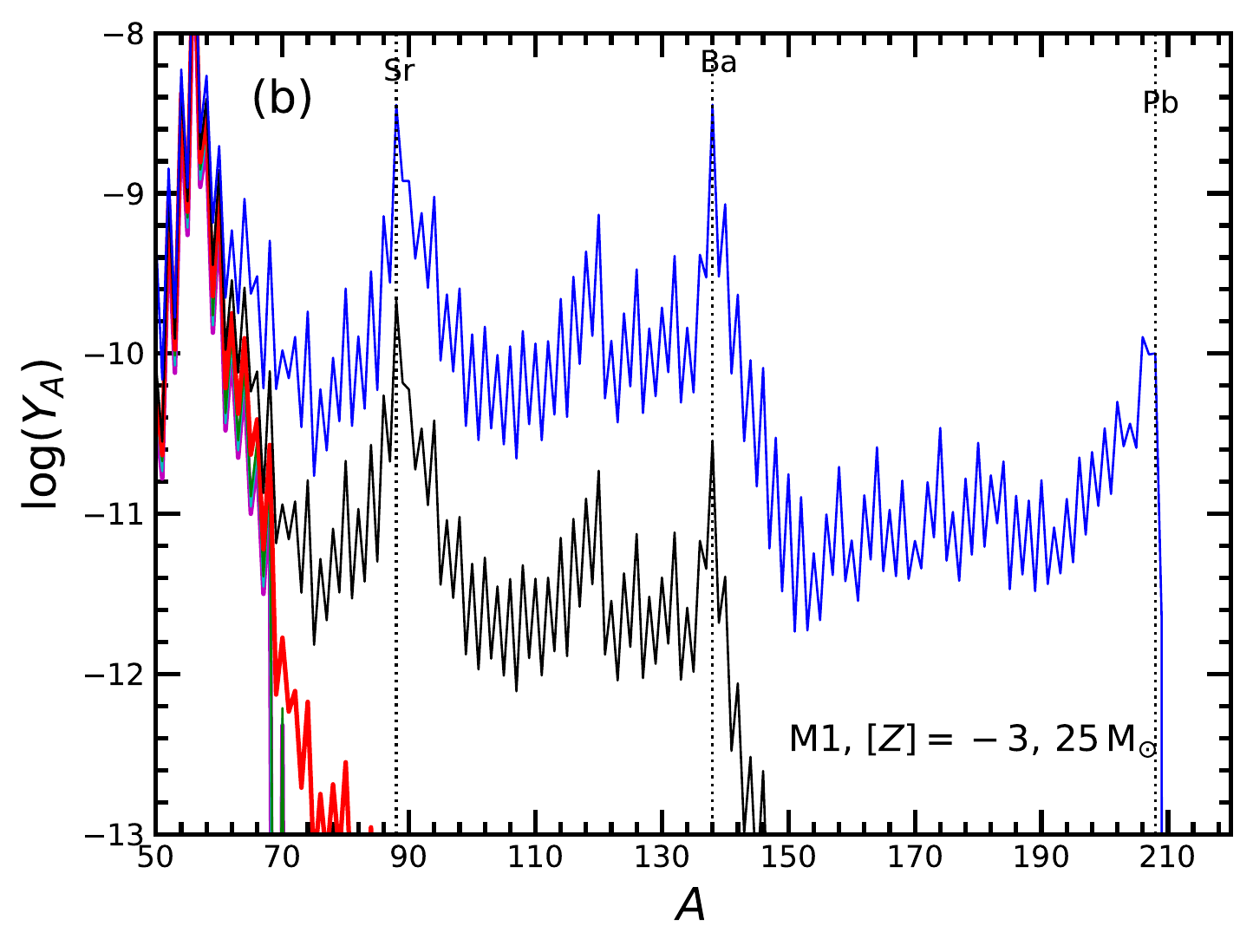}}
\caption{(a) Number yields of heavy isotopes inside the star as functions of mass number corresponding to the evolution stages plotted in Figs.~\ref{fig:snap}a (magenta), \ref{fig:snap}b (cyan), \ref{fig:snap}c (green), \ref{fig:snap}d (red), \ref{fig:snap}e (black), and \ref{fig:snap}f (blue). (b) Same as (a) but for the cumulative wind ejecta at the corresponding stages.}
\label{fig:snapA}
\end{figure}
\subsection{\spr in QCH Models}
We find that all stars that reach the QCH state undergo the \textsl{s}-process. On reaching the QCH state, the star starts to burn He in the center and its core becomes convective. At this stage, although the star is mostly a He star, crucially, it is not completely homogeneous. Specifically, in the outer radiative region of the star, beyond the convective He core, protons are still present, with a typical mass fraction of $\sim 1$--$3\,\%$ (Fig.~\ref{fig:snap}a--c). This turns out to be the key ingredient that enables a subsequent \spr to occur. As the size of the convective He core grows, primary $^{12}$C is synthesized and quickly becomes abundant, with a mass fraction of $\sim0.2$ (Fig.~\ref{fig:snap}c). Due to rotation-induced mixing, some of the $^{12}$C is slowly mixed outwards into the radiative region outside the convective core where protons are available. Because of the QCH evolution, this region has densities of $\sim10$--$30\,\gcc$ and temperatures of $\sim(7$--$8)\times10^7\,\K$ so that the $^{12}$C can react efficiently with the protons. The resulting gradual accumulation of $^{13}$C can reach mass fractions of up to $\sim0.1\,\%$ in this region (Fig.~\ref{fig:snap}c--d). Again due to rotation-induced mixing, some of the $^{13}$C is slowly mixed back into the convective He-burning core. Additionally, some of the $^{13}$C is slowly ingested as the convective core grows outwards. Once the $^{13}$C gets into the convective core, it is transported inward towards hotter regions with a central temperature of $\sim2\times10^8\,\K$, where it can react efficiently with the abundant $^4$He to produce neutrons via the  reaction $^{13}\mathrm{C}(\alpha,\mathrm{n})^{16}\mathrm{O}$. The resulting neutron density is typical of the \textsl{s}-process, ranging from $\sim10^7$ to $10^{8}\,\pcc$ at the center. The corresponding \spr can last the most part of the He-burning phase, which is up to $\sim 10^{13}\,\s$ ($\sim 3\times 10^5$~yr) for a $25\,\Msun$ star. The time evolution of the \spr yield pattern is shown in Fig.~\ref{fig:snapA} for the M1 Model of a $25 \,\Msun$ star with an initial metallicity of $[Z]=-3$ and an initial rotation speed of $\vrot=0.5\,\vc$. Pre-CCSN yields of Sr, Ba, Eu, and Pb are given in Table~\ref{tab:1} for a number of models.

In the same layers where $^{13}$C is synthesized, a substantial amount of primary $^{14}$N is also produced through the CNO cycle (Figs.~\ref{fig:snap}c--f). The $^{14}$N is then mixed into the convective core similarly to $^{13}$C and acts as a potent neutron poison. In addition, during the later stages of core He burning, primarily-produced $^{16}$O becomes sufficiently abundant with a mass fraction of $\gtrsim0.1$ and starts to be an important neutron poison. Near the end of He burning, other primary poisons, such as $^{20}$Ne and $^{24,25,26}$Mg, also begin to act as significant neutron sinks (Figs.~\ref{fig:snap}e--f). Thus, primary neutron poisons become increasingly important to counter neutron production by the primary $^{13}$C during the later stages of core He burning. Nevertheless, neutron capture continues until He is completely exhausted at the center, resulting in an overall strong \spr with production of elements up to Bi.  

It is well known that most of the $^{14}$N mixed into the convective He-burning core is efficiently converted into $^{22}$Ne via $^{14}$N$(\alpha,\gamma)^{18}$F$(e^{+}\nu_{\rm e})^{18}$O$(\alpha,\gamma)^{22}$Ne. $^{22}$Ne can produce neutrons via $^{22}\mathrm{Ne}(\alpha,\mathrm{n})^{25}\mathrm{Mg}$ and is in fact the main neutron source for the rotation-induced \spr at low metallicities found by \citet{pignatari2008} and \citet{frischk2012,frischk2016}. In our QCH models, $^{22}$Ne also becomes the dominant neutron source during the very late stages of He burning (Fig.~\ref{fig:snap}f). Their contribution to the overall strong \textsl{s}-process, however, is negligible compared to that of $^{13}$C.

\begin{table*}
\centering
\vspace{-\baselineskip}
\caption{Pre-CCSN yields (in $\Msun$) of Sr, Ba, Eu, and Pb for selected models reaching the QCH state. Here $X(Y) \equiv X\times10^Y$.}
\vskip 0.5cm
\begin{tabular}{lllll}
\hline
 Model                           &Sr         &Ba     &Eu       &Pb \\
\hline
M1, $[Z]=-3, 0.3\,\vc, 25\,\Msun$ &1.76(-6) &8.11(-10)&1.21(-13)&2.88(-15)\\
Wind                              &6.67(-11) &1.30(-17)&1.92(-27)&3.85(-37)\\
M1, $[Z]=-3, 0.4\,\vc, 25\,\Msun$ &1.93(-6) &1.45(-5)&2.80(-8)&5.98(-6)\\
Wind                              &3.01(-7) &2.74(-7)&1.46(-10)&5.48(-9)\\
M1, $[Z]=-3, 0.5\,\vc, 25\,\Msun$ &1.30(-6) &1.05(-5)&2.86(-8)&1.94(-5)\\
Wind                              &3.62(-7) &6.83(-7)&6.45(-10)&8.95(-8)\\
\hline
M2, $[Z]=-3, 0.3\,\vc, 25\,\Msun$ &1.81(-6) &9.98(-6)&2.30(-8)&2.65(-5)\\
Wind                              &1.05(-10) &1.68(-21)&1.94(-24)&3.11(-22)\\
M2, $[Z]=-3, 0.4\,\vc, 25\,\Msun$ &1.14(-6) &5.63(-6)&1.70(-8)&5.40(-5)\\
Wind                              &1.02(-7) &4.74(-7)&1.05(-9)&8.74(-7)\\
M2, $[Z]=-3, 0.5\,\vc, 25\,\Msun$ &6.70(-7) &1.90(-6)&4.44(-9)&7.61(-5)\\
Wind                              &6.41(-8) &2.40(-7)&5.65(-10)&4.84(-6)\\
\hline
M1, $[Z]=-3, 0.4\,\vc, 15\,\Msun$ &4.40(-6) &1.73(-6)&1.10(-9)&1.69(-8)\\
Wind                              &4.52(-9) &1.17(-11)&3.46(-16)&3.38(-16)\\
M1, $[Z]=-3, 0.5\,\vc, 15\,\Msun$ &1.49(-6) &8.03(-6)&1.19(-8)&3.32(-6)\\
Wind                              &1.56(-7) &6.21(-8)&2.86(-11)&1.27(-9)\\
\hline
M2, $[Z]=-3, 0.4\,\vc, 15\,\Msun$ &1.38(-6) &7.49(-6)&1.37(-8)&8.06(-6)\\
Wind                              &1.64(-8) &1.31(-8)&6.82(-12)&3.86(-10)\\
M2, $[Z]=-3, 0.5\,\vc, 15\,\Msun$ &1.08(-6) &5.07(-6)&1.34(-8)&2.15(-5)\\
Wind                              &7.88(-8) &2.66(-7)&4.63(-10)&2.35(-7)\\
\hline
\end{tabular}
\vspace{-1.0\baselineskip}
\vskip 0.5cm
\label{tab:1}
\end{table*}

\begin{figure}
\centerline{\includegraphics[width=85mm]{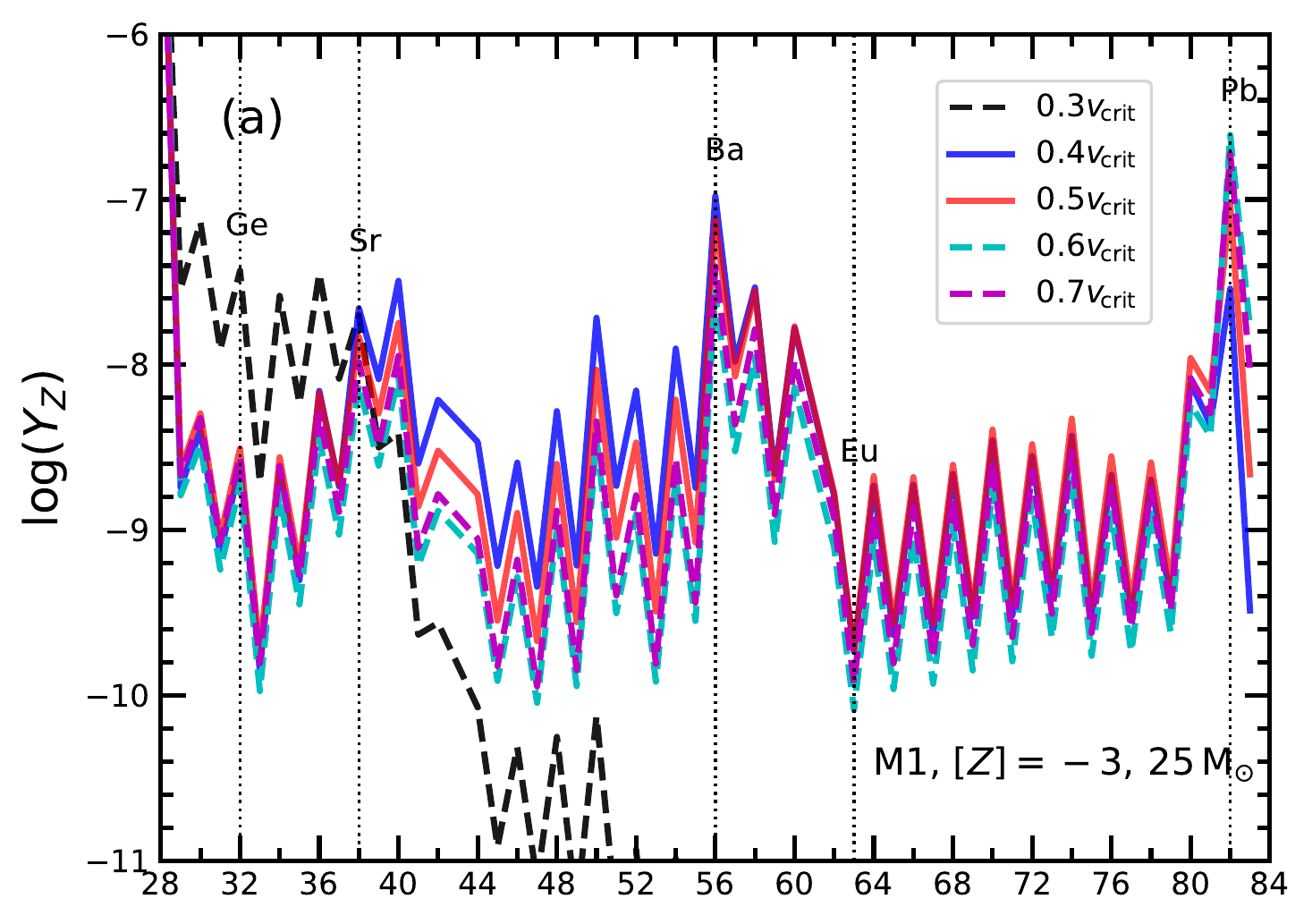}}
\centerline{\includegraphics[width=85mm]{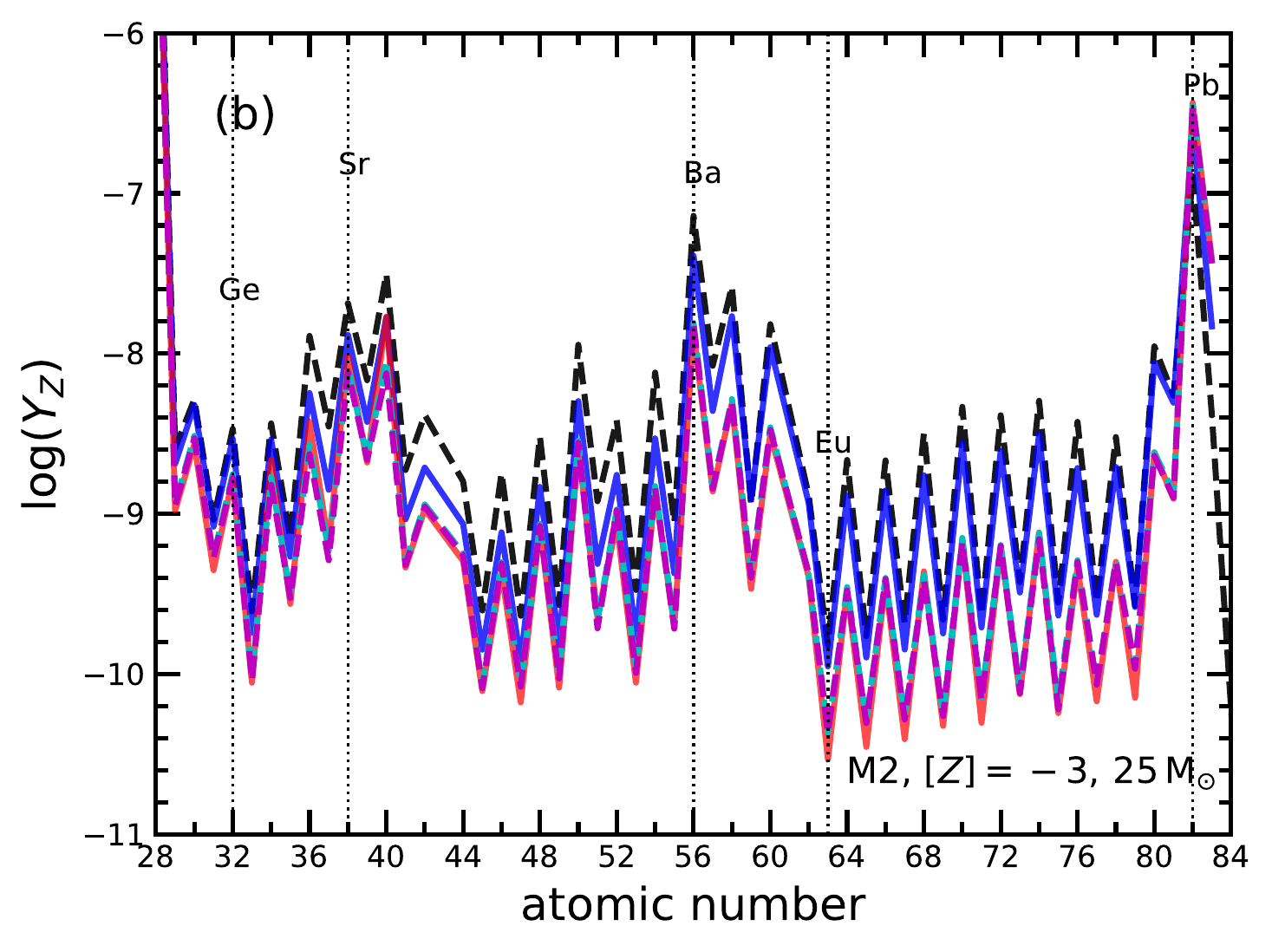}}
\caption{(a) Number yields of heavy elements inside the star as functions of atomic number for the M1 models of a $25\,\Msun$ star with an initial metallicity of $[Z]=-3$ and with varying initial rotation speeds of $30\,\%$ (black), $40\,\%$ (blue), $50\,\%$ (red), $60\,\%$ (cyan), and $70\,\%$ (purple)} of $\vc$. (b) Same as (a) but for the M2 models.
\label{fig:rotmloss}
\end{figure}
\subsection{Effects of Mass Loss and Rotation}
For the same stellar mass, initial metallicity, and initial rotation speed, neutron capture is more efficient in the M2 model with a lower WR mass loss rate, primarily because less mass loss leads to less loss of angular momentum. For a $25\,\Msun$ progenitor with $[Z]=-3$, we find that rotation is faster in the M2 model by $\sim 30\%$ than that in the M1 model during core He burning. Faster rotation induces more mixing, thereby facilitating a more efficient \textsl{s}-process. The effect of mass loss on the \spr efficiency is particularly evident for slower initial rotation speeds near the $v_{\rm min}$ required to reach the QCH state. For example, for a $25\,\Msun$ progenitor with $\vrot=0.3\, \vc$, the \spr flow only reaches Sr ($A\sim 90$) in the M1 model. In contrast, the \spr flow easily reaches Bi ($A\sim 209$) in the M2 model (Figs.~\ref{fig:rotmloss}). 

\begin{figure}
\centerline{\includegraphics[width=85mm]{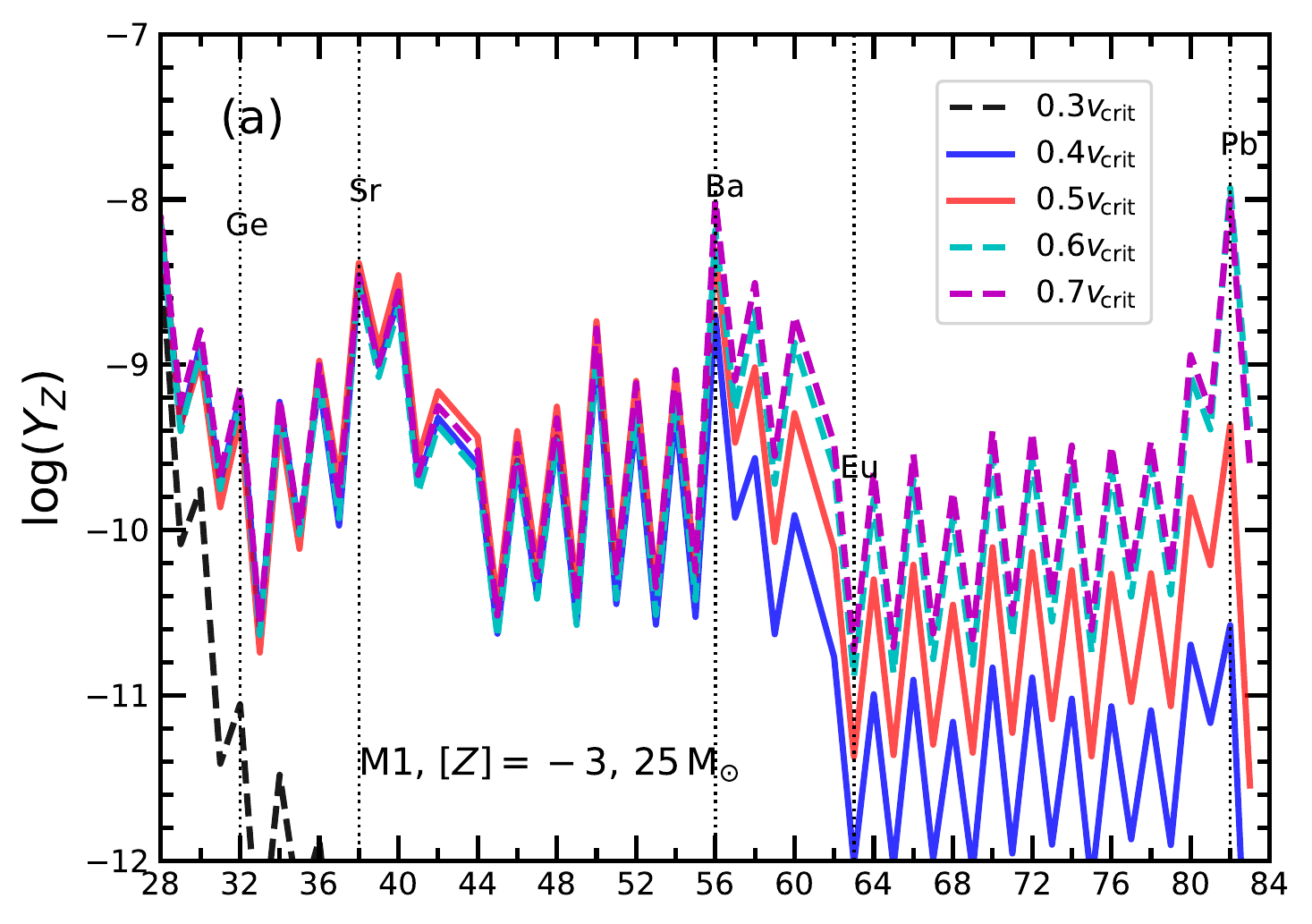}}
\centerline{\includegraphics[width=85mm]{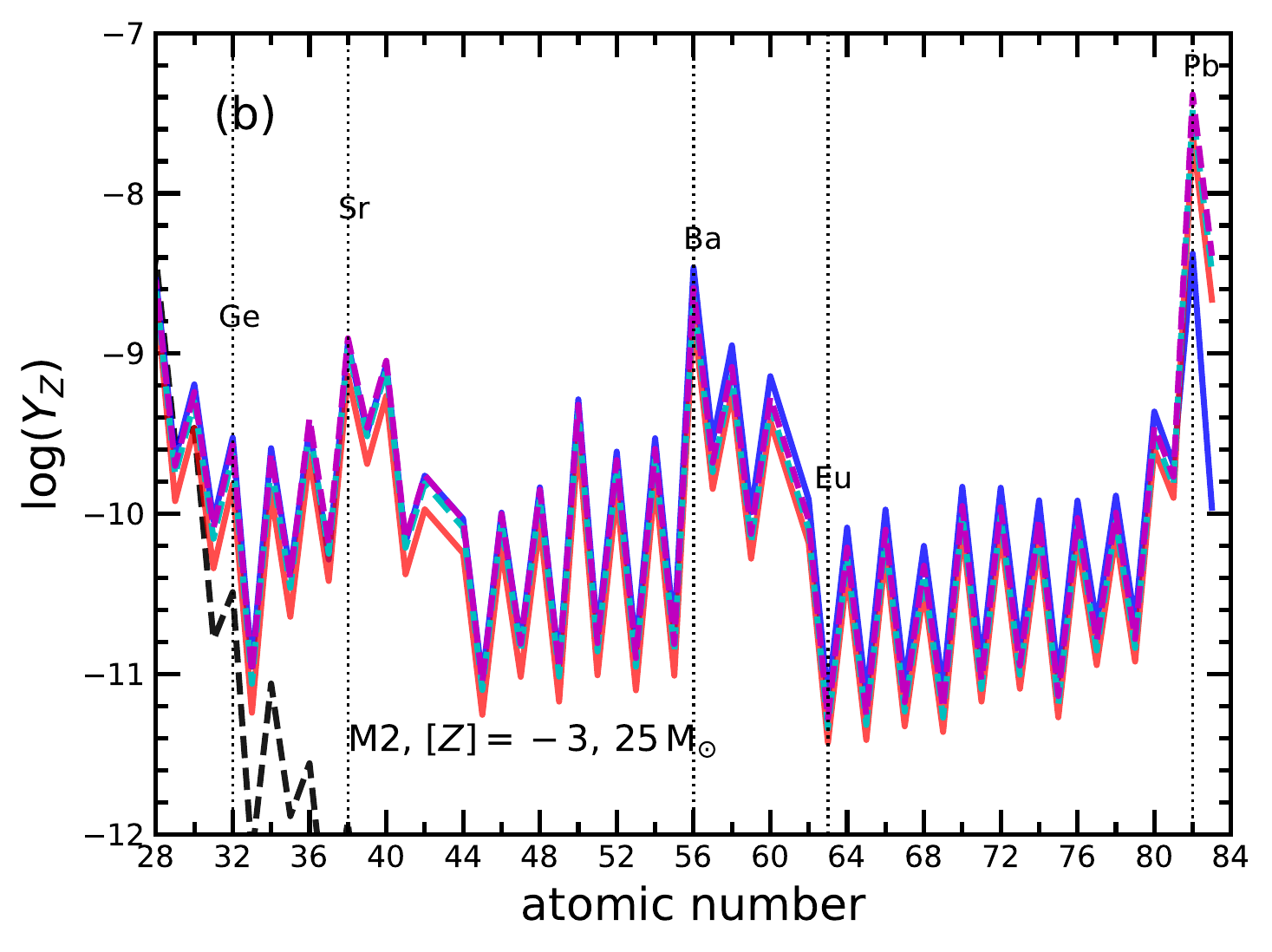}}
\caption{(a) Same as Fig.~\ref{fig:rotmloss}a, but for the wind ejecta. (b) Same as Fig.~\ref{fig:rotmloss}b, but for the wind ejecta.}
\label{fig:rotmlosswind}
\end{figure}

Following the above discussion, we expect that for the same stellar mass, initial metallicity, and mass loss, faster initial rotation would give rise to more efficient neutron capture, and hence, a stronger \textsl{s}-process. This trend, however, is valid only up to a certain rotation speed.  Then the \spr efficiency saturates or even starts to decrease slightly. For example, for the M1 models of a $25\,\Msun$ progenitor with $[Z]=-3$, the \spr efficiency, as indicated by the production of Pb relative to Ba, is marginally less for $\vrot =0.7\, \vc$ than for $\vrot =0.6\, \vc$, whereas the efficiency is almost identical for the corresponding M2 models (Fig.~\ref{fig:rotmloss}-\ref{fig:rotmlosswind}). The saturation, or slight decrease, of the \spr efficiency is caused by the competition between production of the neutron poison, $^{14}$N, and that of the neutron source, $^{13}$C. The former starts to slowly dominate the latter at sufficiently high rotation speed. Nevertheless, the \spr is still very efficient for $\vrot =0.7\, \vc$, the fastest speed considered in our study.

For a $25\,\Msun$ progenitor with $[Z]=-3$, we find that the \spr can robustly produce Bi in the M2 model once the initial rotation is fast enough to reach the QCH state, i.e, $\vrot=0.3\,\vc$. For $\vrot>0.3\,\vc$, a very efficient \spr in the M2 model leads to overproduction of Pb relative to Ba and Sr (Fig.~\ref{fig:rotmloss}b). Clearly, faster initial rotation can compensate for more severe loss of angular momentum due to more mass loss in the M1 model. For a $25\,\Msun$ progenitor, we find that an initial rotation speed of $\vrot \gtrsim 0.4\, \vc$ is required to produce Bi in the M1 model (Figs.~\ref{fig:rotmloss}a). 

As shown in Figs.~\ref{fig:rotmloss} and \ref{fig:rotmlosswind}, the effects of mass loss and rotation on the \spr are evident in the yield patterns both inside the star and for the wind ejecta.

\subsection{Effects of Initial Metallicity}
\begin{figure}
\centerline{\includegraphics[width=85mm]{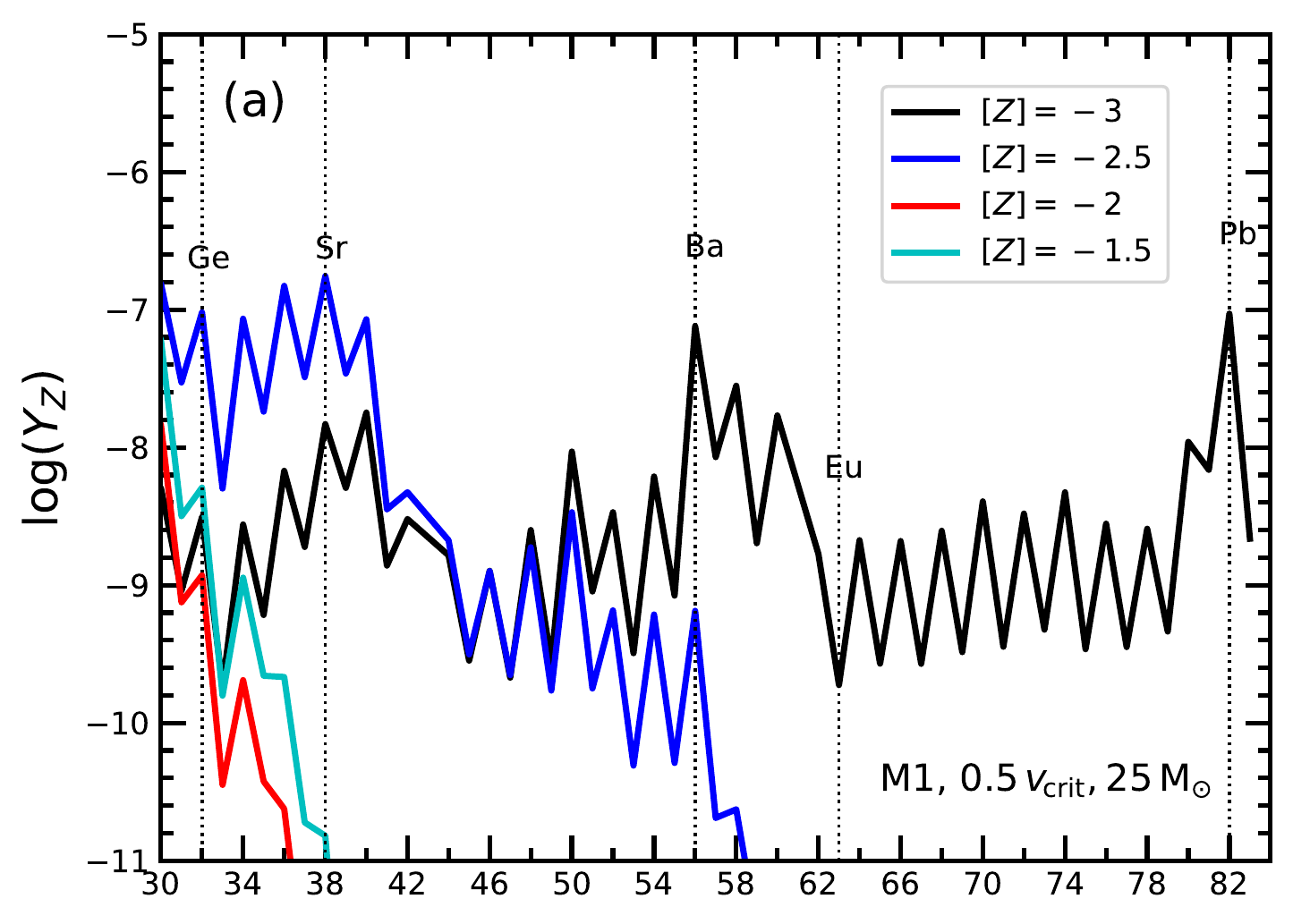}}
\centerline{\includegraphics[width=85mm]{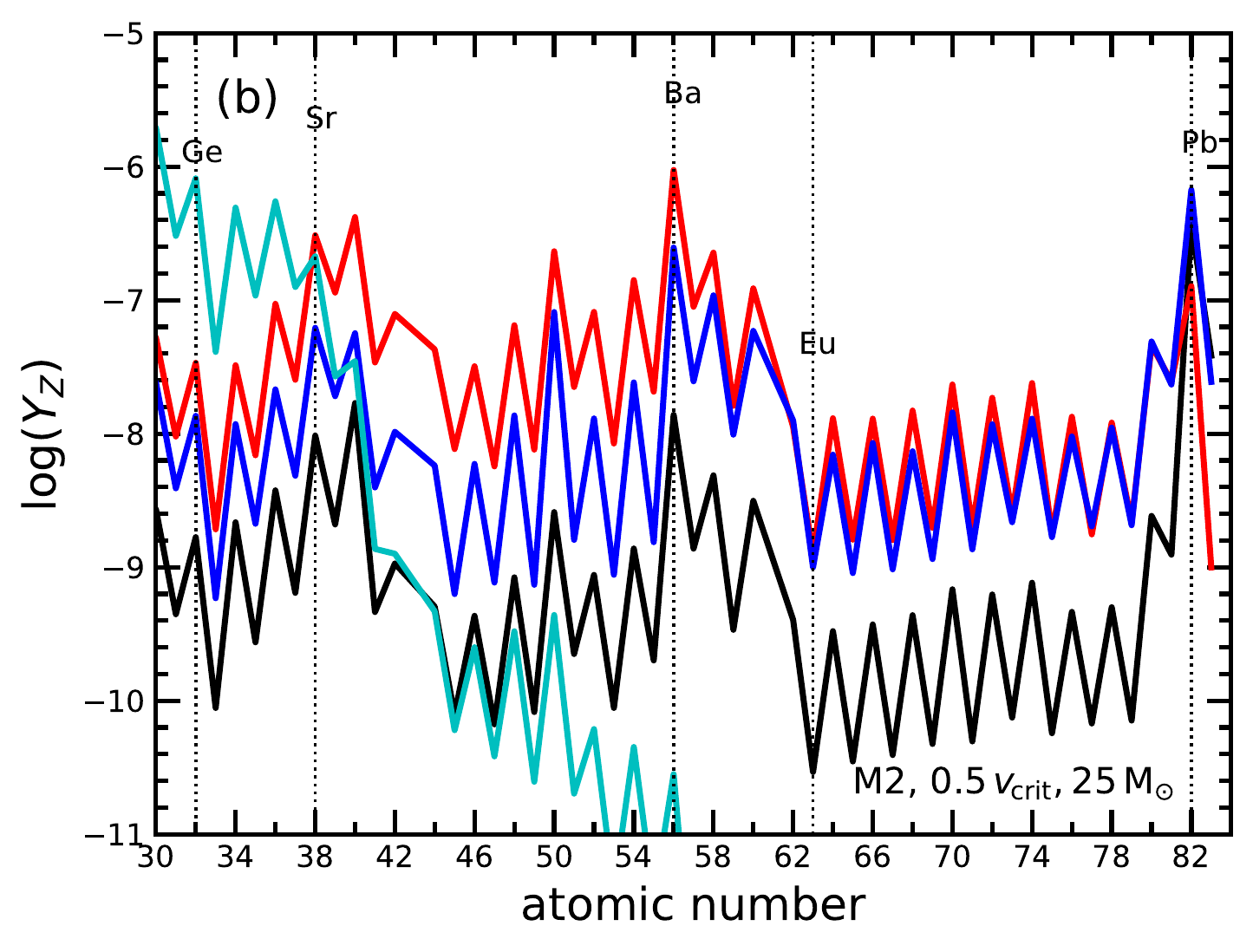}}
\caption{(a) Number yields of heavy elements inside the star as functions of atomic number for the M1 models of a $25\,\Msun$ star with an initial rotation speed of $0.5\,\vc$ and with varying initial metallicities of $[Z]=-3$ (black), $-2.5$ (blue), $-2.0$ (red), and $-1.5$ (cyan). (b) Same as (a) but for the M2 models.}
\label{fig:metallicity}
\end{figure}

As discussed above, mass loss leads to angular momentum loss, which affects rotation-induced mixing responsible for producing the primary $^{13}$C neutron source. Because the mass loss rate increases with the initial metallicity $Z$, the \spr efficiency depends on $Z$. In addition, for progenitors with $[Z]\gtrsim -2$, neutron poisons inherited from the birth material start to become important. Consequently, the efficiency of the \spr is adversely affected in higher-metallicity progenitors, with the effect being particularly strong in the M1 models with more mass loss. For example, for the M1 models of a $25\, \Msun$ progenitor with an initial rotation speed of $\vrot=0.5\,\vc$, rotation during core He burning is a factor of $\sim 2.6$ slower for $[Z]=-2$ than that for $[Z]=-3$, which accounts for the negligible \textsl{s}-processing in the $[Z]=-2$ model (Fig.~\ref{fig:metallicity}a). Similarly, for the M2 models, rotation is slower by a factor of $\sim 1.6$ for $[Z]=-2$ than that for $[Z]=-3$. Rotation in the M2 model for $[Z]=-2$, however, is still $\sim 10\,\%$ faster than that in the M1 model for $[Z]=-3$. Therefore, a substantial \spr occurs in the M2 model for $[Z]=-2$, although it is less efficient, i.e., has a lower Pb/Ba ratio, than the \spr in the M2 model for $[Z]=-3$ (the higher absolute yields for $[Z]=-2$ are due to the more abundant seeds from the higher initial metallicity, Fig.~\ref{fig:metallicity}b). When the metallicity reaches $[Z]=-1.5$, the main \spr is no longer possible, but substantial production of heavy elements up to $A\sim 90$ can still occur through the weak \textsl{s}-process in the M2 models (Fig.~\ref{fig:metallicity}). 

\subsection{Composition of Wind Ejecta}
Because of the QCH evolution, stars in our models reach the WR stage well before H is depleted at the center. Thus, significant mass loss through winds starts long before the occurrence of the \spr during core He burning and continues until the end of the star's life. The composition of the wind ejecta changes over time in terms of both the heavy \spr and the light elements. This change depends on the slow rotation-induced mixing of material from the convective core towards the surface and on the mass loss that affects the efficiency of this mixing. As a result of this mixing, a gradient in composition is produced between the core and the surface. In particular, as the \spr proceeds to heavier elements in the core, the \spr composition on the surface always lags behind that in the core and has a lower enrichment. For example, by the time the \spr in the core has reached Pb, the surface has only been enriched approximately up to Ba (Fig.~\ref{fig:snapA}). Thus, the \spr composition of the wind ejecta may not reach the heaviest elements produced in the core. For the same reason, by the end of the star's life, the net composition of all the wind ejecta is also less enriched in the \spr elements compared to the core (Table~\ref{tab:1}). Nevertheless, the net wind composition can easily reach Bi with substantial enrichment of the heavy \spr elements for the M2 models (Fig.~\ref{fig:rotmlosswind}). In addition to the heavy elements, the wind ejecta is also naturally enriched in CNO from core He burning.      

\subsection{Evolution Beyond Core He Depletion}
\label{sec:ilike}
Although the \spr is essentially over by the end of core He burning, some additional neutron capture can take place during the later stages when the star contracts to burn O at the center. During those stages, the C/O shell still has some He left with a mass fraction of $\gtrsim 10\%$. Due to the overall contraction of the star, the temperature there rises to $\sim 4\times 10^8~\K$ with typical densities of $\sim 10^3~\gcc$ so that He can burn convectively. Whereas $^{13}$C has already been burned away by then, abundant primary $^{22}$Ne with a mass fraction of $\gtrsim 10^{-3}$ in the C/O shell produces neutrons via $^{22}\mathrm{Ne}(\alpha,\mathrm{n})^{25}\mathrm{Mg}$, with the neutron density reaching up to $n_\mathrm{n}\sim 10^{11}\, \pcc$ at the base of the shell. The associated neutron capture can change the abundances of the elements between the \spr peaks and turn the earlier \spr pattern in the C/O shell into an \textit{r/s}-like pattern, i.e, a pattern in between \textit{s}- and \textit{r}-patterns that is usually similar to what is attributed to the intermediate (\textsl{i}) neutron-capture process \citep{cowan1977}. For reference, we note that the solar \textsl{r}- and \spr patterns are characterized by [Ba/Eu]$_\mathrm{r}\sim -0.8$ and  [Ba/Eu]$_\mathrm{s}\sim 1.6$, respectively. In this paper we refer to patterns with $\mathrm{[Ba/Eu]}\gtrsim 1$ as \textsl{s}-like and $\mathrm{[Ba/Eu]}< 1$ as \textsl{r/s}-like.

C burning in the convective O/Ne shell during the later stages of evolution also gives rise to high neutron densities of up to $n_\mathrm{n}\sim 5\times 10^{12}\, \pcc$ at the base of the shell, where the neutron source is again the primary $^{22}$Ne. Initially, neutron capture at higher $n_\mathrm{n}$ at the base of the shell is able to dominate that at much lower $n_\mathrm{n}$ in the outer part of the shell, thereby modifying the abundance pattern in the shell into an \textsl{r/s}-like one. As $^{22}$Ne gets depleted, however, neutron density at the base of the O/Ne shell drops, so that neutron capture at lower $n_\mathrm{n}$ in the outer colder part of the shell brings the pattern back to an \textsl{s}-like one. In contrast, due to the overall low $n_\mathrm{n}$, neutron capture in the outer part of the He burning C/O shell is never able to dominate that at the base of the shell so that the \textsl{r/s}-like pattern is preserved until core collapse. We find that [Ba/Eu] as low as $\sim 0.65$ can be preserved in the  He burning C/O shell.

Although neutron capture after core He depletion can modify the abundance pattern between the peaks as described above, it makes a negligible contribution to the net \spr as the abundance peaks at Sr, Ba, and Pb remain practically unchanged. This result comes about simply because very low neutron exposure is provided during the limited time ($\lesssim 10^8~\s$) available before core collapse. Nevertheless, the higher neutron densities reached during this time can change the \spr pattern into an \textsl{r/s}-like one. By the end of the star's life, however, only the \textsl{r/s}-like pattern in the C/O shell can survive. Nevertheless, it is remarkable that we can obtain an \textsl{r/s}-like pattern by modifying an existing \spr pattern at neutron densities of $\sim 10^{11}$--$10^{12}\, \pcc $, which are much lower than those typically required for an \textsl{i}-process to occur, i.e., $\gtrsim 10^{14}\, \pcc$.

With regard to the structure, the QCH evolution turns the star into a He star, where the He core spans almost the entire star and is comparable to those of non-rotating stars that are $\gtrsim 2$--$3$ times more massive. Following core He depletion and subsequent stages of central C, O, and Si burning in these large He cores, the pre-collapse structure of the $15\, \Msun$ and $25\, \Msun$ models corresponds to that of non-rotating stars of $\gtrsim 40\, \Msun$ with relatively high compactness. A compactness parameter $\xi_M$ \citep{oconnor2011} is defined as
\begin{equation}
    \xi_M=\frac{M/\Msun}{R(M)/1000~{\rm km}},
\end{equation}
where $R_{M}$ is the radius enclosing a certain baryonic mass $M$. The value of $ \xi_M$ just before core collapse is considered a good indicator for the likelihood of a successful neutrino-driven SN explosion, with a low likelihood for $ \xi_M$ above a critical value \citep{oconnor2011, sukhbold2014}. For the usual choice of $M=2.5\, \Msun$, the QCH stars at the pre-SN stage have $\xi_M\sim 0.33$--$0.45$. These values are too large for the standard neutrino-driven explosion to work \citep{ertl2016,muller2016,sukhbold2016}. Therefore, these stars are likely to produce a weak SN or might even fail to explode unless their rapid rotation facilitates non-neutrino-driven explosion such as in magneto-rotational or jet-driven SNe associated with LGRBs (see \citealt{woosley2012} and \citealt{nagataki2018} for reviews) or in SLSNe powered by the spin-down of magnetars \citep{woosley2010,kasen2010}.

Regardless of the details of the explosion, the pre-SN winds from these rotating stars ensure that a substantial amount of the \spr material is always ejected into the interstellar medium (ISM). Any material ejected during the explosion will add to the wind ejecta. Provided that an explosion can be launched, we find that a large amount of material might be ejected even in the presence of severe fallback. For example, with our simple piston prescription for the artificially-induced explosion, the outer $\sim 6$ $(7)\, \Msun$ of the stellar core is ejected in addition to the wind ejecta of $\sim 6$ $(3)\, \Msun$ from a $25\, \Msun$ star with $[Z]=-3$ in the M1 (M2) models for a typical explosion energy of $1.2\times 10^{51}$ ergs. 
Interestingly, the ejecta from the explosion includes the C/O shell with the \textsl{r/s}-like pattern for the heavy elements.
We note, however, that long-term 3D simulations such as those by \citet{chan2018} are required to estimate reliably the amount of material ejected during the explosion. 

\section{Dependence on Treatment of Mixing}
\begin{figure}
\centerline{\includegraphics[width=85mm]{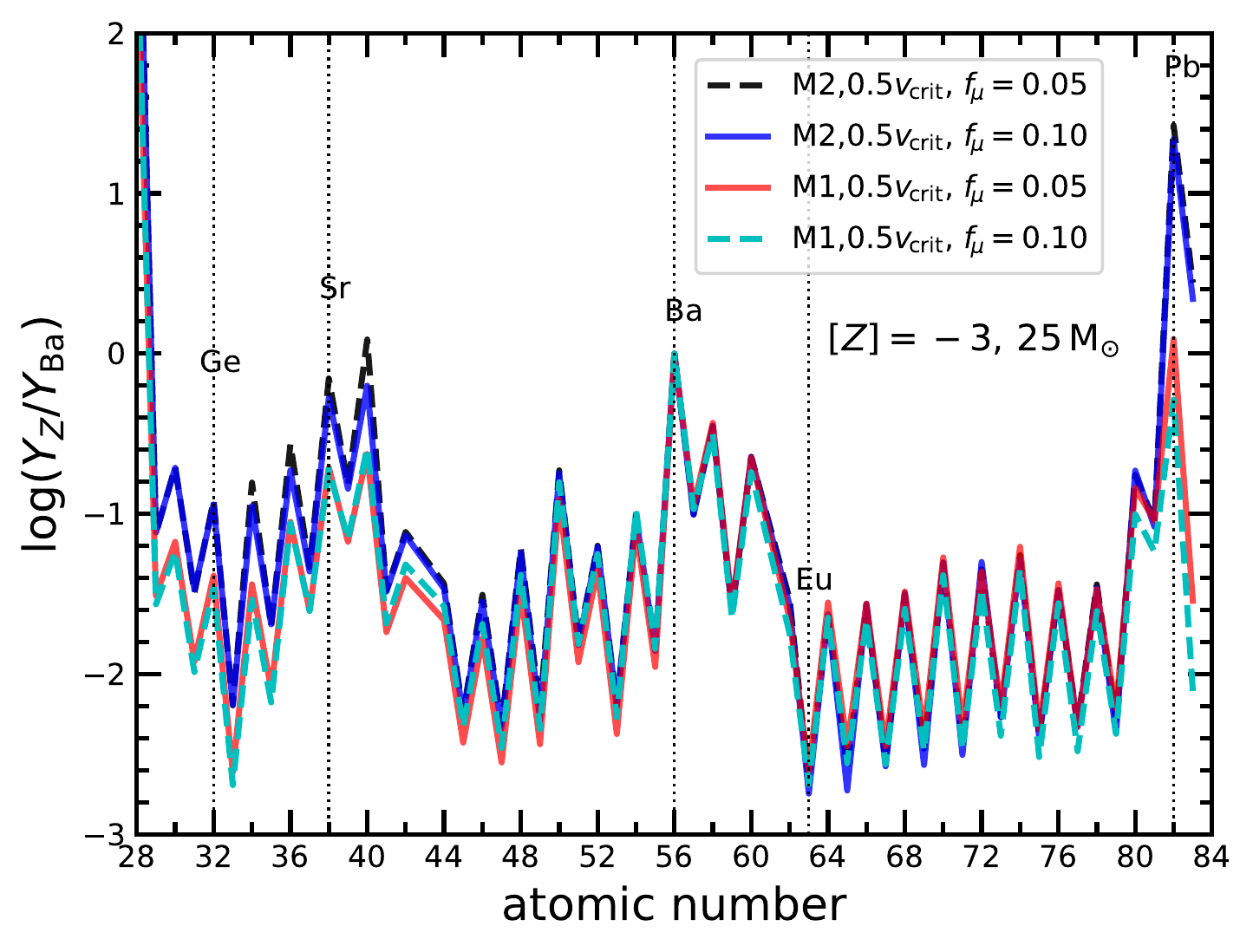}}
\caption{Number yields of heavy elements relative to Ba inside the star as functions of atomic number for the M1 and M2 models of a $25\,\Msun$ star with an initial rotation speed of $0.5\,\vc$ and $[Z]=-3$ for the default value of $\fmu{}=0.05$ compared to the corresponding models with $\fmu{}=0.10$.}
\label{fig:mixtest}
\end{figure}

In order to test the sensitivity of the \spr in QCH models to the particular choice of mixing parameters, we performed another set of calculations for the same $\fc$ but with a twice larger value of $\fmu=0.10$.  For $10$--$20\, \Msun$ stars with typical initial rotational speed of $\sim 200\, \kms$ at solar metallicity, such a choice of parameters gives a slightly lower surface nitrogen enhancement than observed \citep{heger2000a}. For both, the M1 and M2 models of a $25\,\Msun$ star with initial metallicity of $[Z]=-3$, the \textsl{s}-process yields obtained with $\fmu=0.10$ are almost identical to those for the default value of $\fmu=0.05$ (Fig.~\ref{fig:mixtest}). This result suggests that the \spr studied here is not very sensitive to reasonable choice of parameters for rotation-induced mixing.

\begin{figure}
\centerline{\includegraphics[width=85mm]{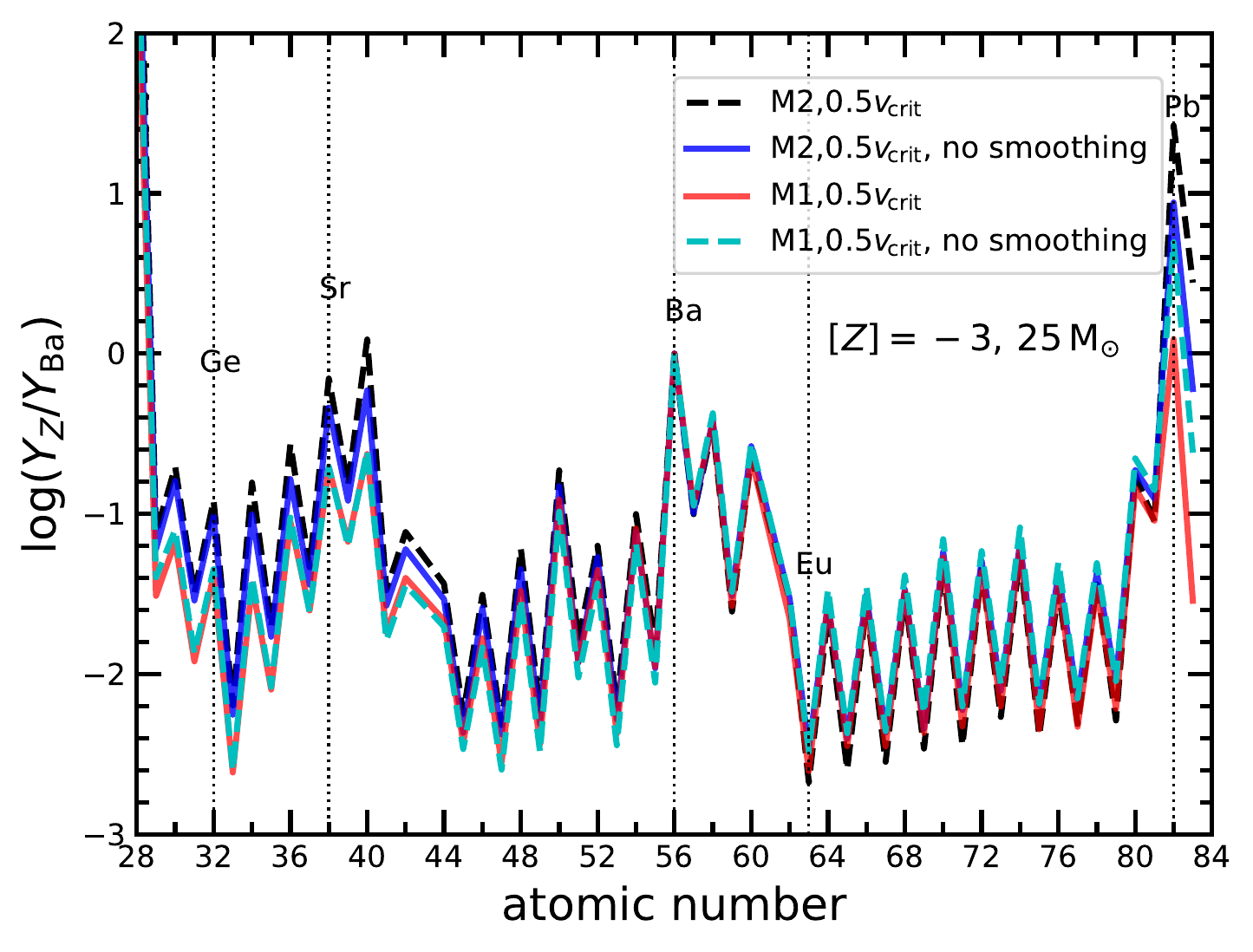}}
\caption{Number yields of heavy elements relative to Ba inside the star as functions of atomic number for the M1 and M2 models of a $25\,\Msun$ star with an initial rotation speed of $0.5\,\vc$ and $[Z]=-3$ using the default smoothing for rotation-induced mixing compared to the corresponding models without smoothing.}
\label{fig:smoothtest}
\end{figure}

As is typical of 1D codes, all our default calculations used smoothing of gradients in treating rotation-induced instabilities. We also repeated the calculations for the M1 and M2 models of a $25\, \Msun$ star with initial metallicity of $[Z]=-3$ without smoothing. The resulting \spr is almost unchanged except for some differences in the Pb abundance when compared to the default calculations (Fig.~\ref{fig:smoothtest}). This indicates that smoothing does not significantly affect the \spr in QCH models.

\section{Discussion and Conclusions}

\begin{figure}
\centerline{\includegraphics[width=85mm]{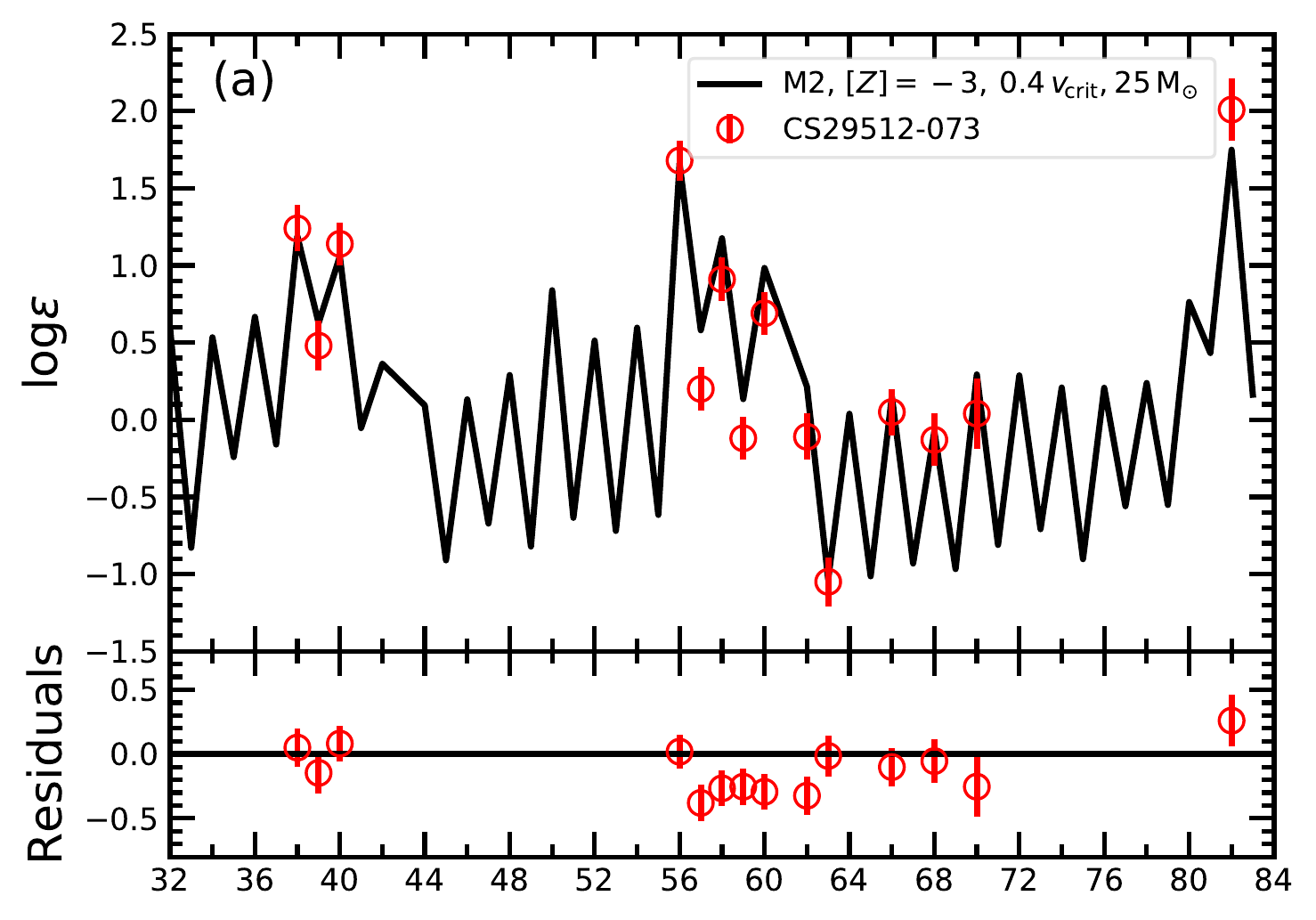}}
\centerline{\includegraphics[width=85mm]{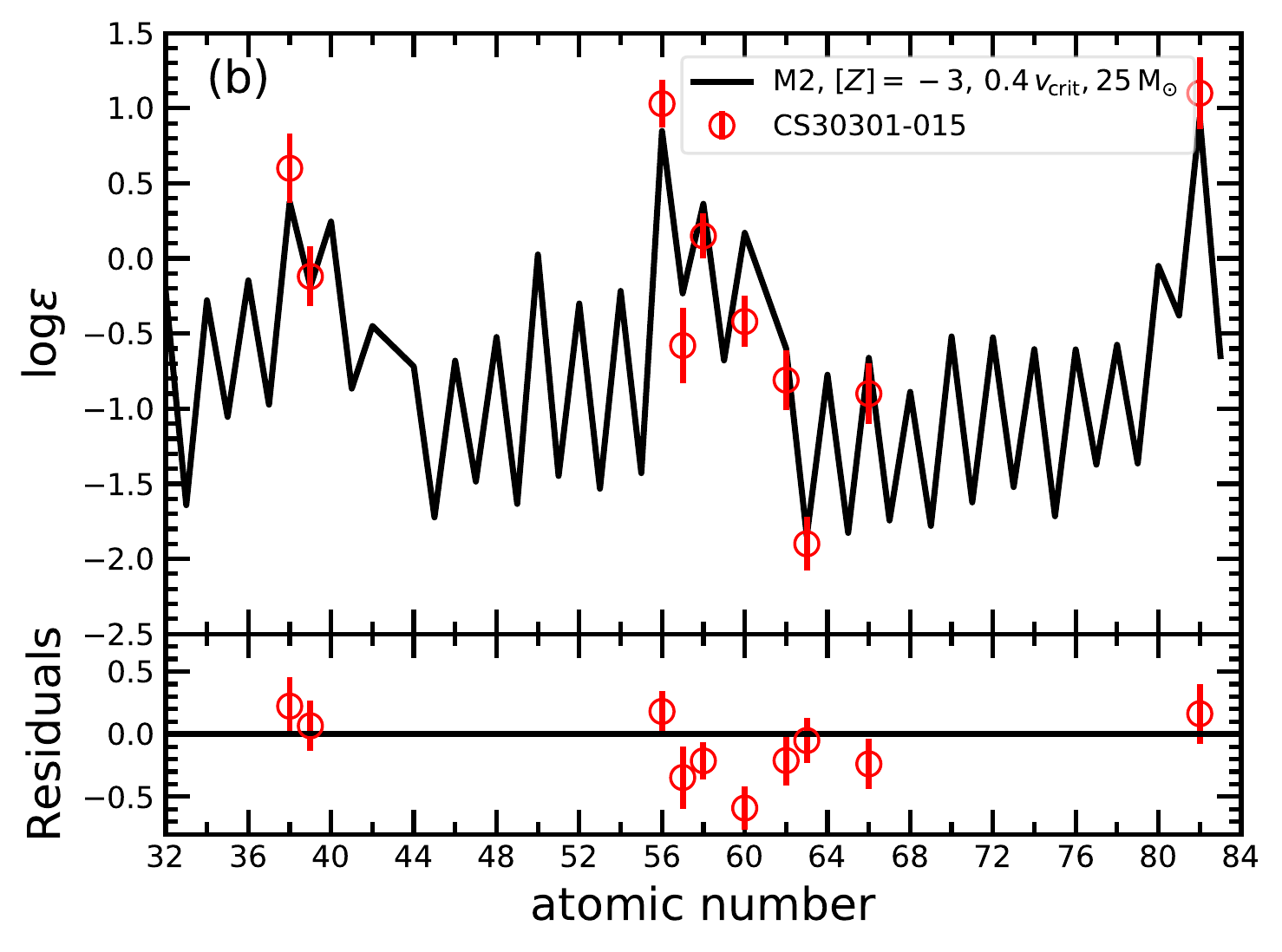}}
\caption{(a) Heavy-element abundances from the wind ejecta in the M2 model of a $25\, \Msun$ star with an initial metallicity of $[Z]=-3$ and an initial rotation speed of $0.4\,\vc$ compared to data for the CEMP-\textsl{s} star CS29512-073 \citep{roederer2014}. A dilution mass of $100\, \Msun$ is used. 
(b) Same as (a), but for the CEMP-\textsl{s} star CS30301-015 \citep{aoki2002} with a dilution mass of $650\, \Msun$.} 
\label{fig:cemps}
\end{figure}

Rapidly-rotating massive metal-poor stars reaching the QCH state can be a natural site for the main \textsl{s}-process. Without becoming fully chemically homogeneous He stars, these QCH stars have a relatively minor amount of H present in the outer parts, which ultimately gives rise to the $^{13}$C neutron source through rotation-induced mixing. Although neutron production in these stars is independent of the initial metallicity, the efficiency of the \spr decreases with increasing metallicity. Higher metallicity leads to more mass loss, which in turn slows down the rotation, thereby decreasing the efficiency of rotation-induced mixing critical to the \textsl{s}-process. A strong \spr is possible only in QCH models of massive stars with initial metallicities of $[Z]\lesssim -1.5$. 

The \spr in rapidly-rotating massive metal-poor stars discussed here would have direct implications for the chemical evolution of heavy elements in the early Galaxy as revealed by observations of elemental abundances in low-mass ($\lesssim 0.8\, \Msun$) very metal-poor (VMP) stars with [Fe/H]~$\lesssim -2$. Such VMP stars are thought to have formed within $\sim 1$ Gyr after the Big Bang and their surface abundances are considered to directly reflect the composition of the ISM from which they were formed. Because the main \spr is usually associated with long-lived low-mass stars that could not have contributed to the chemical enrichment during the first $\sim 1$ Gyr, it is commonly assumed that the heavy elements with $A\gtrsim 120$ observed in VMP stars must have originated exclusively from the \textsl{r}-process associated with early short-lived massive stars. Yet deviations from the expected \textsl{r}-process pattern, which is characterized by [Ba/Eu]$_\mathrm{r}=-0.8$, were observed in VMP stars with [Fe/H] as low as $\sim -3.5$ \citep{saga,simmerer}. Further, there exists a sub-class of VMP stars called the carbon-enhanced metal-poor (CEMP) stars \citep{beers2005}, which are highly enriched in C with $\log \epsilon({\rm C})\sim 8.0 \pm 0.4$) and [C/Fe]~$>0.7$ \citep{spite2013,bonifacio2015,hansen2015,yoonCEMPno1}. Most of the CEMP stars are the so-called CEMP-\textsl{s} stars, which are also enriched in heavy elements with [Ba/Fe]~$>1$ and have abundance patterns exhibiting [Ba/Eu]~$>0.5$, similar to what is expected from the main \textsl{s}-process. The \spr discussed here can explain at least some of the above observations.

Clearly, our models of rapidly-rotating massive metal-poor stars provide an early source for the main \spr that naturally accounts for the early onset of this process as revealed by the abundance patterns of some VMP stars. In addition, our models can explain at least some CEMP-\textsl{s} stars. Whereas most CEMP-\textsl{s} stars are in binaries \citep{lucatello2005}, observations suggest that $\sim 10$--$30\,\%$ of them could be single stars \citep{hansenCEMPs}. If a rapidly-rotating massive metal-poor star had a low-mass binary companion, transfer of the wind material ejected by the massive star onto the surface of the companion would have provided high enrichment of C and the heavy \spr elements to the latter. It is possible that the binary would have survived the subsequent evolution and explosion of the massive star, especially if the star collapsed directly into a black hole or the explosion imparted little kick to the black hole left behind. In this scenario, some binary companions of CEMP-\textsl{s} stars would be black holes of $\sim 10\,\Msun$ instead of white dwarfs. It is also likely, however, that the binary would have been disrupted. In this case, the low-mass star with its surface already polluted would be observed as a single CEMP-\textsl{s} star today. Of course, single CEMP-\textsl{s} stars could also have been formed directly from the ISM enriched by the winds of rapidly-rotating massive metal-poor stars. In any case, the heavy elements in the wind ejecta can provide good fits to the observed \spr patterns. Figure \ref{fig:cemps} shows example fits to the data on two CEMP-\textsl{s} stars using the wind ejecta from the M2 model of a $25\, \Msun$ star with $[Z]=-3$ and $\vrot=0.4\,\vc$. 

A popular model for explaining the high C and \spr enrichment in CEMP-\textsl{s} stars involves surface pollution due to mass transfer from a binary companion during its AGB phase (see e.g., \citealt{bisterzo2}). Although this model can successfully explain the abundance patterns in many CEMP-\textsl{s} stars, it has difficulty fitting the patterns in other such stars \citep{bisterzo3,abate2015}. As it naturally predicts that all CEMP-\textsl{s} stars should be accompanied by white dwarfs left behind by their former ABG companions, it is also in apparent conflict with the existence of single CEMP-\textsl{s} stars. As discussed above, the \spr in rapidly-rotating massive metal-poor stars can account for at least some CEMP-\textsl{s} stars, including those in single configuration. Therefore, this explanation is complementary to the popular model for CEMP-\textsl{s} stars.

\begin{figure}
\centerline{\includegraphics[width=85mm]{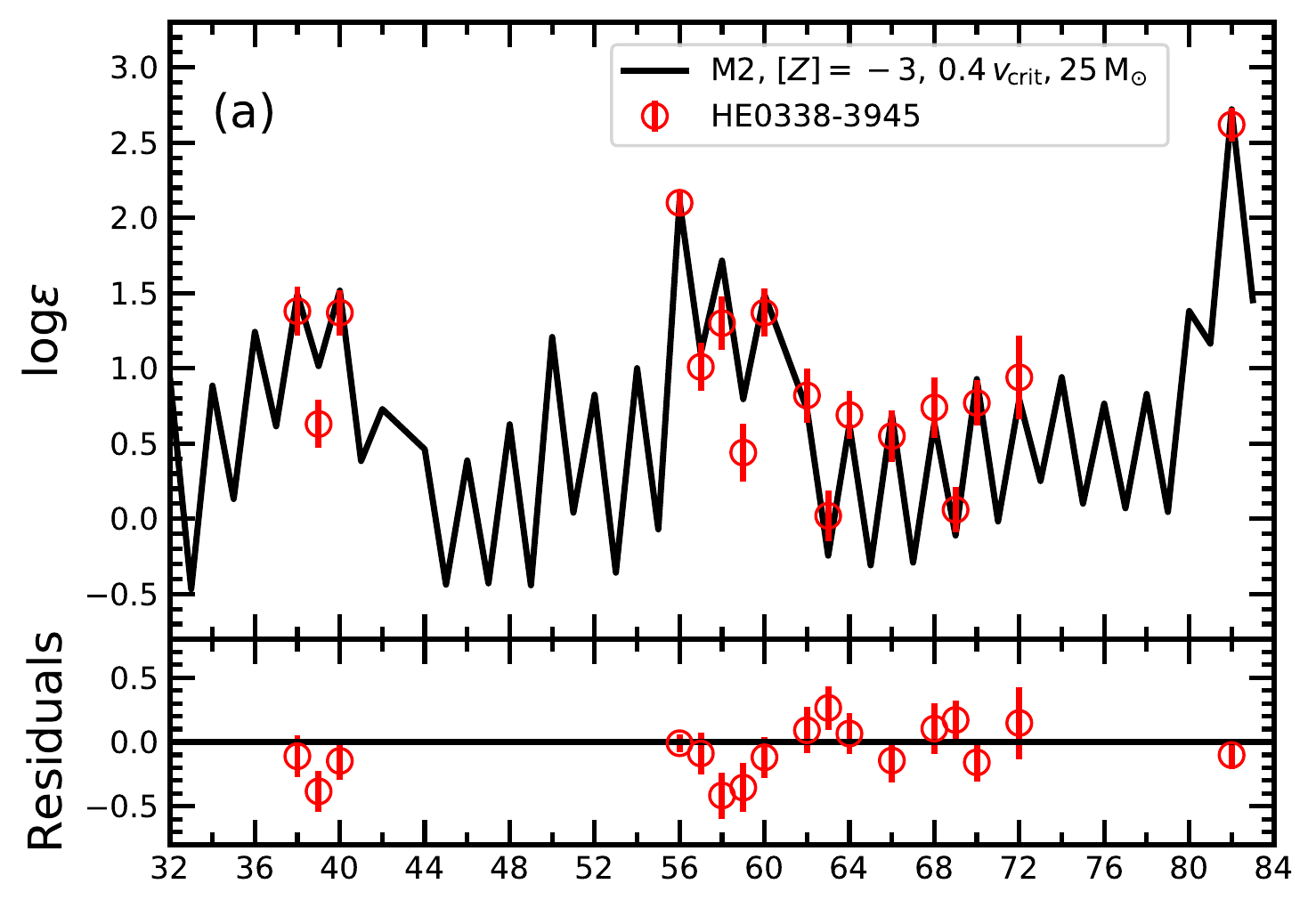}}
\centerline{\includegraphics[width=85mm]{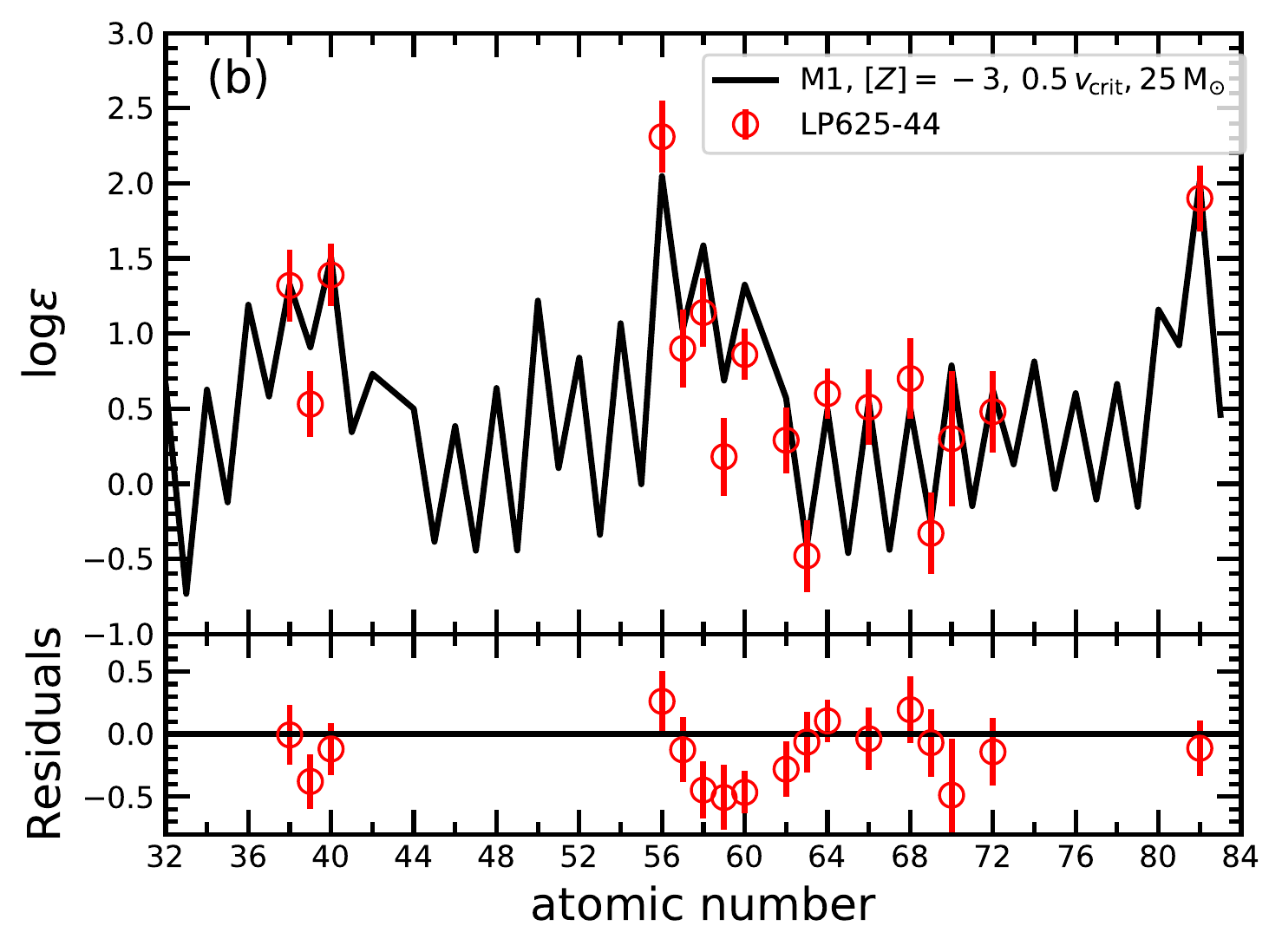}}
\caption{ (a) Heavy-element abundances from the mixture of the outer stellar and wind ejecta in the M2 model of a $25\, \Msun$ star with an initial metallicity of $[Z]=-3$ and an initial rotation speed of $0.4\,\vc$ compared to data for the CEMP-\textsl{r/s} star HE0338-3945 \citep{jonsell2006}. A dilution mass of $180\, \Msun$ is used.  (b) Same as (a), but for the M1 model with an initial rotation speed of $0.5\,\vc$ compared to data for the CEMP-\textsl{r/s} star LP625-44 \citep{aoki2002b} with a dilution mass of $180\, \Msun$.}
\label{fig:cemprs}
\end{figure}

CEMP stars also include the so-called CEMP-\textsl{r/s} stars. These stars have similar properties to CEMP-\textsl{s} stars, i.e, high enrichment in C and heavy elements, but have abundance patterns that appear to be a mixture of those from the \textsl{r}-process and the \textsl{s}-process. The origin of the heavy elements in CEMP-\textsl{r/s} stars is still unresolved, although a number of models have been proposed \citep{cohen2003,barbuy2005,jonsell2006,dardelet2014}. Recent studies showed that the \textsl{i}-process might be the most likely explanation \citep{lugaro2012}. This process can naturally produce the abundance patterns in most CEMP-\textsl{r/s} stars \citep{hampel2016}. Consequently, these stars are now also referred to as CEMP-\textsl{i} stars. The proposed sites for the \textsl{i}-process include low to intermediate mass stars of $\lesssim 8\, \Msun$ \citep{fujimoto2000,campbell2010,herwig2011,jones2016}, massive stars of $\sim 20$--$30\, \Msun$ \citep{bqh_pingest} or more \citep{clarkson2017}, and accreting white dwarfs \citep{deni2017,deni2019}.
In our models, during core O burning, the outer part of the star, specifically the C/O shell, experiences additional neutron capture that eventually modifies the abundance pattern from \textsl{s}-like to \textsl{r/s}-like. If this material is ejected in addition to the wind ejecta, the resulting abundance pattern can fit some of the CEMP-\textsl{r/s} stars. Figure \ref{fig:cemprs} shows example fits to the data on two CEMP-\textsl{r/s} stars using the mixture of the outer stellar and wind ejecta from a $25\, \Msun$ star with $[Z]=-3$ for the M2 model (with $\vrot=0.4\,\vc$) and M1 model (with  $\vrot=0.5\,\vc$), respectively. For the model shown in Fig.~\ref{fig:cemprs}a, adding the outer stellar core material to the wind changes [Ba/Eu] from 1.03 to 0.69 while for the model in Fig.~\ref{fig:cemprs}b, [Ba/Eu] changes from 1.65 to 0.88. Surprisingly, we can explain these two CEMP-\textsl{r/s} stars with our models despite our neutron densities are much lower than those  required for the \textsl{i}-process ($\gtrsim 10^{14}\, \pcc$).  As discussed in \S~\ref{sec:ilike}, an \textsl{r/s}-like pattern is obtained in our models by modifying an earlier \spr pattern at neutron densities of $\sim 10^{11}$--$10^{12}\, \pcc$. For this reason, we do not refer to CEMP-\textsl{r/s} stars as CEMP-\textsl{i} stars here. The lowest [Ba/Eu] that can be achieved in the outer stellar core, however, is $\sim 0.65$, hence CEMP-\textsl{r/s} with $\mathrm{[Ba/Eu]}< 0.65$ cannot be explained by the rotating models discussed here. The \textsl{i}-process might be required to explain such stars, making them bona fide CEMP-\textsl{i} stars. 

Another site for neutron capture in non-rotating massive metal-poor stars \citep{bqh_pingest} can also help explain the early onset of the \textsl{s}-process, CEMP-\textsl{s} stars, and CEMP-\textsl{r/s} stars. This site relies on proton ingestion into convective He shells to provide neutrons and can produce a variety of abundance patterns including those similar to what are attributed to the main \textsl{s}-process and the \textsl{i}-process. Whereas 3D models are required to calculate reliably the ingestion and transport of protons, only parametric 1D studies were performed by \cite{bqh_pingest} to explore the associated nucleosynthesis. Ideally, 3D models should also be used to follow the evolution of rotating stars, especially rotation-induced mixing that is critical to neutron production in the 1D QCH models presented here. We are well aware of the limitations of our 1D models, but note that they are guided by observations and have been used to study topics other than nucleosynthesis.

This work covers a relatively small grid of progenitor mass, initial metallicity, and initial rotation speed. The limited set of models studied, however, is sufficient to elucidate the mechanism and main features of the \spr in rapidly-rotating massive metal-poor stars, which is the main goal of this paper. We plan to calculate a larger set of models to obtain a more comprehensive picture. In particular, the dependence on the two free parameters \fc{} and \fmu, which control the efficiency of rotation-induced mixing, needs to be explored further.  We stress, however, that their values adopted in this work are identical to those used in earlier studies and were not tuned to obtain the \textsl{s}-process of interest here. In this sense, the \spr is likely robust in rapidly-rotating massive metal-poor stars reaching the QCH state. Nevertheless, whether a star can reach the QCH state may depend on the physics model and numerical implementation.  Whereas the QCH state in rotating massive stars was reported by \citet{scyoon2005}, \citet{scyoon2006}, and \citet{woosley2006}, other studies did not find this state \citep{pignatari2008,frischk2012,frischk2016,choplin2018}.  The latter make different choices with regard to use of magnetic fields and dynamo, use different smoothing algorithms, etc.  Although the treatment of mixing in all these models was calibrated to match the observed surface $^{14}$N enhancement at core hydrogen depletion for massive stars with $\vrot \sim 200$--$300\, \kms$ at solar metallicity, they give different results for the evolution beyond core hydrogen depletion and for faster rotation speeds that are not constrained by observations.  Clearly, these differences are due to those in the use and implementation of rotation-induced mixing in 1D stellar evolution codes.  Future code comparison studies are required to shed light on this issue. Finally, as shown here, mass loss impacts the efficiency of the \spr in QCH models by slowing down rotation. In view of the uncertainties in mass loss from WR stars, especially at very low metallicities, the effects of different mass loss prescriptions also need to be explored further. In addition, the synthesis of CNO and other light elements in QCH models will be explored in the near future.

\acknowledgements
This work was supported in part by the National Natural Science Foundation of China [11533006 (SJTU), 11655002 (TDLI)], 
the US Department of Energy [DE-FG02-87ER40328 (UM)], and the Science and Technology Commission of Shanghai Municipality [16DZ2260200 (TDLI)].

\software{\textsc{Kepler}} \citep{weaver1978, rauscher2003,heger2005}.


\end{document}